\documentclass{emulateapj}
\usepackage[symbol]{footmisc}
\usepackage{longtable}
\usepackage[export]{adjustbox}
\def\lap{\lower.5ex\hbox{$\; \buildrel < \over \sim \;$}}
\def\gap{\lower.5ex\hbox{$\; \buildrel > \over \sim \;$}}

\def\ergcm2s{${\rm erg\ cm^{-2}\ s^{-1}}$}

\def\ergscm2s{${\rm erg\ cm^{-2}\  s^{-1}}$}

\def\cm-2{${\rm cm^{-2}}$}

\def\Msun{${\rm{\,M_\odot}}$}
\def\pc{{\rm{\,pc}}}

\def\Mpc{{\rm{\,Mpc}}}
\def\Myr{{\rm{\,Myr}}}

\def\mzams{M$_{\rm ZAMS}$}
\def\pasa{PASA}
\begin{document}

\renewcommand{\thefootnote}{\fnsymbol{footnote}}
\title{Constraints for the Progenitor Masses of Historic Core-Collapse Supernovae}

\author{Benjamin F. Williams\altaffilmark{1},
Tristan J. Hillis\altaffilmark{1},
Jeremiah W. Murphy\altaffilmark{2},
Karoline Gilbert\altaffilmark{1,3},
Julianne J. Dalcanton\altaffilmark{1},
Andrew E. Dolphin\altaffilmark{4}
}
\altaffiltext{1}{Department of Astronomy, Box 351580, University of
  Washington, Seattle, WA 98195; ben@astro.washington.edu;
  tristan3214@live.com; jd@astro.washington.edu}
\altaffiltext{2}{Department of Physics, Florida State University, jeremiah@physics.fsu.edu}
\altaffiltext{3}{Space Telescope Science Institute; kgilbert@stsci.edu}
\altaffiltext{4}{Raytheon, 1151 E. Hermans Road, Tucson, AZ 85706; dolphin@raytheon.com}


\begin{abstract}

We age-date the stellar populations associated with 12 historic nearby
core-collapse supernovae (CCSNe) and 2 supernova impostors, and from
these ages, we infer their initial masses and associated
uncertainties.  To do this, we have obtained new {\it HST} imaging
covering these CCSNe.  Using these images, we measure resolved stellar
photometry for the stars surrounding the locations of the SNe.  We
then fit the color-magnitude distributions of this photometry with
stellar evolution models to determine the ages of any young existing
populations present.  From these age distributions, we infer the most
likely progenitor mass for all of the SNe in our sample.  We find ages
between 4 and 50 Myr, corresponding to masses from 7.5 to 59 solar
masses. There were no SNe that lacked a young population within 50 pc.
Our sample contains 4 type Ib/c SNe; their masses have a wide range of
values, suggesting that the progenitors of stripped-envelope SNe are
binary systems. Both impostors have masses constrained to be $\lap$7.5
solar masses. In cases with precursor imaging measurements, we find
that age-dating and precursor imaging give consistent progenitor
masses.  This consistency implies that, although the uncertainties for
each technique are significantly different, the results of both are
reliable to the measured uncertainties.  We combine these new
measurements with those from our previous work and find that the
distribution of 25 core-collapse SNe progenitor masses is consistent
with a standard Salpeter power-law mass function, no upper mass
cutoff, and an assumed minimum mass for core-collapse of
7.5~M$_{\odot}$.  The distribution is consistent with a minimum mass
$<$9.5~M$_{\odot}$.

\end{abstract}

\section{Introduction}

Supernovae play a key role in nucleosynthesis, the formation of
neutron stars and black holes, and in the morphological and chemical
evolution of galaxies.  The quantitative outcomes of these phenomena
depends on which massive stars actually explode, and so it is
important to constrain the distribution of massive stars that explode.
However, their progenitors are difficult to study because supernovae
are rare.  Therefore, while there are well-developed theories of
stellar evolution leading up to supernovae, these theories are
difficult to test with observations that link SNe to their progenitor
stars.

Stellar-evolution theory for isolated massive stars makes two basic
predictions for core-collapse supernovae (CCSNe).  First, the
zero-age-main-sequence mass (\mzams) and mass-loss history control
whether a SN occurs, and second, the \mzams\ determines the type of
CCSN \citep{woosley02,heger03a,dessart11}.  This theory predicts that
the lower mass stars will explode with their hydrogen envelopes intact
(e.g. II-P, II-L, IIn), and that the most massive stars lose much of
their envelopes and explode as hydrogen deficient CCSNe (e.g. IIb,
Ib/c).  

Given the complexity of the underlying physics, however, especially binary
evolution, winds, and episodic eruptions, it is unclear whether nature
obeys the same well-delineated mass-dependence.  In fact, recent
theory \citep{claeys11,dessart11} and the relatively high observed
rates of H-deficient CCSNe \citep{smith11c} imply that binary
evolution may figure prominently in producing the H-deficient
CCSNe.

One sensitive test of theoretical SNe explosion mechanisms is the
distribution of initial stellar masses that produce
CCSNe. Unfortunately, the number of available progenitor masses of
CCSNe with known spectral types is still relatively small, with each
measurement having significance
\citep[e.g.,][]{smartt09b,maund11,murphy11a,vandyk2017a,davies2017}.
At this point, about 32 CCSNe have progenitor mass constraints from
direct precursor imaging, with many more having only upper bounds
\citep[e.g.,][]{smartt09b,galyam09,smith11b,maund11,murphy11a,fraser2012,vandyk2012a,maund2013,fraser2014,vandyk2013,vandyk2014b,smartt2015,davies2017,vandyk2017}.

To increase the number of progenitor mass measurements available, we
have been using a robust technique for constraining the progenitor
ages for nearby historical CCSNe.  We can measure these ages using
resolved photometry of adjacent stellar populations from {\it Hubble Space
Telescope (HST)} imaging
taken at any time before or after the event (e.g.,
\citealp{badenes2009,gogarten09a,murphy11a,williams2014}, hereafter
  W14; \citealp{maund2017}).  Our method
relies on the fact that only relatively massive stars
($\gap$7.5~M$_{\odot}$) become CCSNe \citep{jennings2012}.  Thus, for
single stars, their lifetimes are limited to $\lap$50~Myr
\citep{girardi2002}. Over 90\% of stars form in clusters containing
more than 100 members with $M_{\rm cluster}\!>\!50$~M$_{\odot}$
\citep{lada03}.  These stars remain spatially correlated on physical
scales up to $\sim$100~pc during the $100\Myr$ lifetimes of
4~M$_{\odot}$ stars, even if the cluster is not gravitationally bound
\citep{bastian06}; we have confirmed this expectation empirically in
several cases \citep[][W14]{gogarten09a,murphy11a}.  Thus, it is
reasonable to assume that any CCSN is physically associated with other
young stars of the same age. This technique has significantly
increased the sample of nearby SNe progenitor mass constraints for
CCSNe with spectral types \citep{williams2014,maund2017}.

Together, the progenitor masses measured to date suggest that the
maximum mass for SN II-P may be lower than expected \citep{smartt09a}.
Counter to expectations, some H-rich CCSNe (in particular IIn) have
been associated with more massive stars
\citep{galyam09,smith11b,smith11c,fox2017,dwek2017}. Furthermore, no progenitor mass of
$>$20~M$_{\odot}$ has been established for any historical SNe with
certainty, hinting at an upper mass cutoff [W14].  However, there are
recent indications that no such maximum mass exists
\citep[e.g.,][]{davies2017}, and may be the result of biases in
analysis or sample size.  Additional measurements are needed to
significantly improve tests of any such cutoff.

We recently obtained new observations with {\it HST} of 12 more nearby
CCSNe and 2 supernova impostors with known spectral types, allowing us
to increase the number of color-magnitude diagram (CMD)-based constraints further.  With this
paper, the CMD-based sample is now up to 25 SNe (not
including impostors), which we use to place constraints on the
progenitor mass distribution.

In this paper, we describe our analysis of these new data and provide
interpretations of the results in the context of massive stellar
evolution and core-collapse supernova theory.  Section 2 describes the
data obtained and provides a brief summary of our analysis technique,
which is already described in detail in previous papers.  Section 3
details the results for our new constraints.  Section 4 merges these
results with previous measurements to investigate the shape
of the progenitor mass distribution function, and section 5 concludes
with a summary of our findings.

\section{Data Acquisition and Analysis}

Here we detail our sample selection, observations taken, resolved
stellar photometry measurements, and stellar evolution model fitting.  

\subsection{Sample Selection}

 We selected all historical CCSNe within $8 \Mpc$ with core-collapse
 spectral types that also have modest foreground extinction and
 arcsecond accuracy in their positions (Table~\ref{sample}).  {\bf Such
 accuracy should be sufficient given the size of the apertures within
 which the CMD is constructed.}  We then cross-referenced the targets
 with the {\sl HST} archive and identified those CCSNe that did not
 yet have adequate {\sl HST} imaging, which are the 14 listed in
 Table~\ref{sample}.

 The distances in the table are updated to include only {\bf tip of the red
 giant branch (TRGB)} distances, which are the most appropriate to use
 for CMD fitting because it minimizes systematic offsets between
   the stellar evolution models and CMD features.  Two of the SNe
 were in NGC~628, which had several distance measurements within 8~Mpc
 when we first developed the sample.  However, the TRGB distance is 10
 Mpc \citep{jang2014}.  This difference of 0.5 mag in depth results in
 us only reaching absolute magnitudes of -2 instead of -1.5 in these
 cases, which barely reaches the main-sequence turn-off for a 50 Myr
 population instead of reaching below it.  We detected very few stars
 for those SNe, but we were still able to put constraints on the
 progenitors because those stars align well to a narrow range in young
 age.  Thus, our technique appears to provide constrains even out to
 10 Mpc.

Two other SNe (SN~2002bu and SN~2008S) might be SN
 impostors.  We include these SN impostors because their
 classification is not definitive and because it is possible that
 these transients are also associated with the last stages of stellar
 evolution, in which case our proposed method to derive progenitor
 masses is still useful.

\subsection{Observations}

We provide a summary of our observations in Table~\ref{obs_table}, all
from our program GO-14786. We
observed 3 pointings in NGC~6946, covering 7 SNe, 1 pointing in
NGC~628, covering 2 SNe, 1 pointing in NGC~4242 covering 1 SN, 1
pointing in NGC~1313 covering 1 SN, 1 pointing in NGC~4945 covering 1
SN, 1 pointing in NGC~4618 covering 1 SN, and 1 pointing in NGC~4214
covering 1 SN.  We observed with ACS/WFC and WFC3/UVIS in parallel to
provide the largest possible coverage of the host galaxies.  For two
of our pointings (NGC~628 and NGC~6946-3), this strategy allowed us to
cover additional SNe locations in a single pointing.  For others, it
simply added coverage of these galaxies to the archive.

The exposure times and filters were selected to provide strong age
constraints for the ages we are interested in ($<$50 Myr).  These
measurements require precise photometry of the main sequence stars
produced at these ages.  Our goal was to reach below the main-sequence
turnoff for a 50~Myr population, which is M$_B\,\sim$-2.5 and
M$_V\,\sim$-2.2, corresponding to m$_B$=28.3 and m$_V$=28.2 in
NGC~6946, our most distant (m-M=28.9) and high extinction
(E$_{B-V}$=0.34) target.  We therefore have longer exposure times for
our more distant target pointings.  Our exposure dates, durations, and
depths are provided in Table~\ref{obs_table}.  We note that the
observation we used for SN~2017eaw was a parallel exposure that was not
intended for this program when we planned the observation.  It is
shallower than the other data, but serendipitously it turned out to be
useful for constraining the progenitor mass of SN~2017eaw both from
population analysis and precursor analysis.

\subsection{Photometry}

The photometry for this project was all performed using the DOLPHOT
package \citep[updated HSTphot][]{dolphin2000,dolphin2016}, through
the same photometry pipeline used for the Panchromatic Hubble
Andromeda Treasury, as described in \citet{williams2014a}.  There have
been a few updates to the photometry routine which we outline here.
First, we now use the images that have been corrected for charge
transfer efficiency effects ({\tt flc} images) as inputs for our
photometry instead of correcting our photometry catalogs for CTE
effect after point spread function fitting.  Second, we now use the
TinyTim PSF libraries in DOLPHOT instead of the Anderson libraries, as
rigorous testing found the TinyTim libraries to provide better spatial
uniformity in DOLPHOT.  

One key assumption of our technique is that the young populations
  within $\sim$50 pc of a SN location is coeval with the progenitor
  itself.  Stellar cluster studies suggest that over 90\% of stars
form in clusters containing more than 100 members with
$M\!>\!50$\Msun \citep{lada03}.  Furthermore, these stars likely remain spatially correlated on physical scales up to
$\sim\!100\pc$ during the $100\Myr$.  This spatial correlation
continues even for low mass clusters that are not gravitationally
bound \citep{bastian06}.  Our previous studies
\citep{gogarten09a,murphy11a} have confirmed that this assumption is
reasonable.  We show some example CMDs of the data for the
full field with the SNe location (within 50 pc) photometry overlaid in
Figures~\ref{example1} and \ref{example2}.  All of these are included as supplemental data for
the article.

Using the same package, we also inserted and recovered artificial
stars in the same regions as those extracted for population analysis
around the locations of the SNe.  These tests are performed by adding
a star with known properties to the original images, and then
rerunning the photometry on the images to determine if the star is
recovered and how close the measured magnitude is to the true
magnitude.  By running many thousands of such tests within the region
of the images containing the SNe locations, we can accurately model
the photometric bias, error, and completeness as a function of color
and magnitude.  All of these quantities are required to fit the
color-magnitude diagrams with those generated from models to derive
the age distribution of the population, as described below.

\subsection{Model Fitting}

We fit all of our SNe locations using the software package MATCH 2.7
\citep{dolphin02,dolphin2012,dolphin13}.  This package has been
well-tested, and has been used in many studies for determining age
distributions of stellar populations in external galaxies
\citep[e.g.,][and many
  others]{dolphin2003,gallart2005,williams2009,weisz2011,hillis2016,skillman2017}. We
have also used it in many previous studies of progenitor mass
estimates
\citep{gogarten09a,jennings2012,jennings2014,williams2014,murphy11a}.

In brief, the MATCH package generates model CMDs for a grid of ages
and metallicities using the stellar evolution models of
  \citet{girardi2002}, updated by \citet{marigo2008}, and
  \citet{girardi2010}. We limited the recent star formation
  metallicities to be from one third solar to solar, as the upper main
  sequence feature is not very sensitive to metallicity, and the
  present day metallicities of these galaxies are not expected to be
  less than the Large Magellanic Cloud.  In translating age to
  progenitor mass, this uncertainty in metallicity can make a
  difference of $<$10\%.

The MATCH software convolves the model grid CMDs with the bias,
uncertainty, and completeness measured with the artificial stars tests
on the data.  It applies extinction prescriptions supplied by the
user, and finds the combination of the model CMD grid plus an optional
background CMD that best fits the observed CMD of the region, where
the weighting of the models corresponds to star formation rates for
the various points in the age-metallicity grid.

The prescription for deriving extinction requires care.  Each SNe
likely resides in a region that is extincted by local dust, which
causes differential extinction that spreads the photometry along the
reddening vector of the CMD, as well as foreground dust that moves the
entire CMD along the reddening line. To quantify these effects, we run
a grid of model fits covering a range of overall extinction and
differential extinction parameters.  We then find the values for these
parameters that provide the best fit to the data, and use those values
for our final fit to the data.  In most cases, the exact value of
these parameters has little impact on the final age distribution.  We
show some examples of the impact of changing the extinction parameters
in Figures~\ref{example1} and \ref{example2}.

The optional background CMD is important in galaxies with a large
amount of global star formation, as is the case with several of our
galaxies.  In this case, we use the CMD of the entire field, which
covers a much larger area of the galaxy.  This background is then
scaled by the ratio of the area of the SNe region to the area of the
field (typically $<$0.001).  In this way, if there is a widespread
population with an age of, for example, 30 Myr, it will be
down-weighted by the amount of stellar mass expected in the surrounding
galaxy.  If there is star formation at this age at a higher density
than the background, it will still be the best-fit age; however, if
there is star formation at this age, as well as at another age, the
other age will be given a higher probability because of its absence
from the global galaxy population.  Use of this background CMD is a
new addition to our technique, and it has helped to isolate the local
population more clearly in some cases. 

Once the best-fit is determined, the package has routines ({\tt
  hybridMC} and {\tt zcombine}) for performing Monte Carlo simulations
and processing them to assess the uncertainties in the star formation
rates for each age \citep{dolphin2013}.  With this information about
the star formation rates, we can determine the relative amount of
stars present at each age.  To measure this relative fraction, we
study the cumulative distribution of the stellar mass with age.
Because of the large amount of covariance between adjacent time bins
in our technique (if the star formation increases in one time bin, it
must decrease in the neighboring time bin for the total model to
remain similar), the uncertainties on the star formation in each
individual time bin tend to be large, making it difficult to
interpret.  However, these highly covariant uncertainties often result
in a relatively clean cumulative distribution which is simpler to
analyze.

We assume that only stars younger than 50 Myr are potential CCSN
progenitors.  This assumption means that we will only measure
progenitor masses $>$7.3~M$_{\odot}$, which is reasonable given the
minimum mass of 7.5~M$_{\odot}$ measured by \citet{jennings2012}.  We
first determine the total stellar mass present that is younger than
this age.  We then calculate the fraction of this total that is
present at each age from 4 to 50 Myr, and assign this as the
probability that the progenitor had that age.  In most cases, there is
a significant peak in the age distribution that allows us to constrain
the age reliably.  This age is only measured for the surrounding
stars, and is therefore not dependent on the detailed evolution of the
system itself.  This age should therefore not be affected by, for
example, the binarity of the progenitor or the uncertain physics of
the last stages of stellar evolution.

Once we have the age measurement, we infer the mass of the progenitor
using model stellar lifetimes
\citep{girardi2002,marigo2008,girardi2010} as a function of stellar
mass.  For this step, we assume the progenitor was a single star. If
the progenitor was in a mass-exchange binary, then the conversion from
age to mass is less clear.  Mass exchange can cause stars to evolve on
different timescales than predicted from their initial masses because,
for example, if a star that is initially 8~M$_{\odot}$ gains
5~M$_{\odot}$ in mass from a close companion, it may end its main
sequence lifetime in significantly less time than an isolated
8~M$_{\odot}$ star.  Thus, for such a system, our age to mass mapping
will not provide a correct progenitor mass, but the progenitor age
should still be useful for modeling the evolution of the progenitor.
Whatever did explode, binary or single, it reached the end of its
stellar lifetime at the age we have measured, so that if it is found
to have been a binary, any model of the binary's evolution would need
to reproduce our measured lifetime.

\section{Results}

We show examples of the star formation histories used to determine the
SN progenitor ages and masses in Figures~\ref{example1} and
\ref{example2}.  Similar plots for all of the SNe are supplied in the
article supplemental data.  Table~\ref{mass_table} provides the best
age and mass with uncertainties for each SN, and Table~\ref{age_dist}
gives the full age probability distribution for each SN for more
rigorous statistical combinations.

{The measurements for the older SNe may be somewhat less reliable due
  to the lower precision astrometry available for those events.
  However, as long as the astrometry is accurate to $\sim$1$"$, our
  extraction will contain the SN location, and the population will be
  representative.  The fact that the measurements for these do not
  appear to be outliers in the mass distribution provides some
  assurance that our extraction regions for these events are
  appropriate.  In one particular case (SN~1954A), we aligned the SN
  from the chart of \citet{Pietra1955} to a 2MASS image of NGC~4214,
  and found that the position was consistent with that of
  \citet{lennarz2012}, which is the position we list is
  Table~\ref{sample} to within an arcsec.  We measured the age both at
  this location and the default location from the Open Supernova
  Catalog (which lists multiple reported locations).  We found that
  while the median age changed {\bf substantially, it remained consistent
  within the large uncertainty} for this SN.  The new position has a
  high median mass, but a lower limit on the mass of $>$11
  M$_{\odot}$.

It is important to note that the median age is not always the best
representation of the mass probability, which is why we include the
full probability distributions for each SN.  Some of the SNe were in
regions where more than one age was clearly present.  Examples of
these are SN~1948B and SN~1968D, which both had very young populations
present of 10 Myr or younger, but these did not contain enough stellar
mass to become the median age.  In these cases, there is a possibility
that the progenitor was younger ($8-10$ Myr) and more massive (20-30
M$_{\odot}$) which is only properly accounted for by using the full
probability distribution we have included in Table~\ref{age_dist}.

For SNe with measurements here and in W14 (SN~1917A, SN~2004et, and
SN~2005af) we prefer the measurement made here. The archival data are
relatively poor in these cases (they had 7 stars or less in W14).  For
example, the previous measurement of SN~2005af was based on only 6
stars and found mass of 9$^{+5}_{-1}$~M$_{\odot}$; however, with our
new data we have 39 stars and a good background measurement, which
argues for a 8~Myr old population above the background and pushes the
best-fit mass to 38$^{+1}_{-27}$~M$_{\odot}$.  While the best fit
changed significantly, the uncertainties of the measurements still
overlap, as that slightly older population is still there at lower
significance.   SN~1917A went from 7
stars to 49 stars, and the mass went from 7.9$^{+6.7}_{-0.5}$ to
15$^{+1}_{-5}$.  The measurement for SN~2004et went from 6 stars to 9
stars and the resulting mass estimate was equivalent to that in W14.
Thus the measurements here are preferred to those of W14 for these SNe
which had much shallower data prior to our new observations.

In one case (SN~2002bu) no young population was detected, suggesting a
mass $<$7.3~M$_{\odot}$.  Since this event was one of the likely
impostors, this result is consistent with that possibility.  The other
possible impostor (SN~2008S) did have a young population present, but
it has the lowest mass estimate of any of the SNe, which is also
consistent with it being an impostor, since the mass estimate is below
the standard limit for core-collapse progenitors. These results are
consistent with previous work suggesting that SN~2002bu and SN~2008S
are the same class of transient with relatively low mass progenitors
\citep{szczygiel2012}.

\section{Discussion}

Excluding the two impostors, we combine our new results with the other
13 non-impostor SNe from W14 and \citet{murphy11a} to study the mass
distribution.  We compared these measurements to nine measurements
from direct precursor imaging from the literature (\citealp[SN
  1987A,][]{woosley88}; \citealp[SN~1993J,][]{stancliffe2009};
\citealp[SN~2003gd,][]{hendry2005}; \citealp[SN~2004dj,][]{wang2005};
  \citealp[SN~2004et,][]{li2005}; \citealp[SN~2005cs,][]{maund2014b};
  \citealp[SN~2008bk,][]{maund2014}; \citealp[SN~2011dh,][]{maund2011};
  \citealp[SN~2017eaw][]{vandyk2017}).  We show a comparison to these measurements in
  Figure~\ref{compare}.  There are other measurements that we could also compare, but we are not providing a
  complete review here.  As examples, \citet{vandyk2002} constrained the mass of the
  SN~1993J progenitor;  \citet{crockett2011} provided another measurement
  of the progenitor mass of SN~2004et.  \citet{li2006} also provide an
  estimate of the SN~2005cs progenitor mass. \citet{vandyk2012} provide a
  measurement of the SN~2008bk progenitor.  There are also very recent measurements of progenitor
  masses in \citet{davies2017}.  

 Similar to W14, we find that the
  population-derived masses are consistent with the precursor imaging
  masses where both are available.  The largest outlier is SN~2004et,
  which has a very young best-fit age by our technique.  The upper
  limit on age is large enough that a significantly lower mass,
  similar to that found for the precursor, is allowed; however, if the
  best-fit age is correct, it would suggest the precursor colors were
  not well-modeled.  For example, either poorly constrained pre-SN
  evolution, bolometric correction issues, or binary evolution could
  make the precursor colors difficult to interpret
  \citep{fuller2017,davies2017,podsiadlowski1992}.  Overall, the
  sample size and uncertainties do not show any
  statistically-significant offset between the measurement techniques,
  and they rule out an offset of greater than $\sim$1~M$_{\odot}$.

Our lowest mass constraints are for the 2 supernova impostors in our
sample.  The two impostors in W14 (SN~1954J and SN~2002kg) were not
the lowest of the sample, but neither were above 13~M$_{\odot}$.  These therefore may be a different class of transient that results
  from more massive progenitors. The photometry and spectra of
  \citet{humphreys2017} suggest that at least one of these (SN~2002kg)
  has a very massive (60$-$80~M$_{\odot}$) progenitor, which is
  somewhat at odds with surrounding populations.

The mass constraints from the surrounding populations should help us
to shed light on their origins. We find that SN~1954J and SN~2002kg
appear to have a different, more massive, progenitor type from
SN~2002bu and SN~2008S, but {\bf not as massive as} 50~M$_{\odot}$ (W14).
  There is a tendency to conflate some SN Impostors with luminous blue
  variables \citep[LBVs, see][for a review]{smith2011b}.  However,
  before our mass constraints there was some indication that SN 2002bu and SN 2008S may represent a lower mass class \citep{szczygiel2012,adams2016}. We note that even the stellar population ages associated with the more massive impostors, SN 1954J and SN 2002kg, are at odds with standard mass inferences for LBVs.  If we assume single-star evolution, then the inferred masses for these impostors are $<$25~M$_{\odot}$.  In contrast, the luminosity inferred mass for LBVs is 50 M$_{\odot}$ and greater \citep{smith2011b,humphreys2017}.

More recently, \citet{smith2015} noted that LBVs are isolated from other massive O stars and suggested that their evolution is dominated by binary evolution.  \citet{aghakhanloo2017} modeled this isolation and suggested that LBVs could be rejuvenated stars that started out as 20 M$_{\odot}$ and through binary evolution became more massive and more luminous.  If this model is correct, the stellar populations surrounding LBVs should be significantly older than the luminosity-derived mass would predict.  The impostors SN 1954J and SN 2002kg show exactly this discrepancy.  From the luminosity, \citet{humphreys2017} infer a mass of 60-80 M$_{\odot}$ for SN~2002kg.  We find that the stellar ages are much older than the lifetime of such a massive star, leading to the lower inferred mass.  LBVs and the most massive SN Impostors may be rejuvenated stars resulting from close binary evolution.  

In Figure~\ref{mass_dist} we plot the mass distribution, color-coded
by SN type, for all 25 CCSNe with population-derived progenitor mass
estimates from this study and W14 (after removing the 2 likely
impostors from this study and 2 more in W14).  The complete
distribution is consistent with standard stellar initial mass
functions (IMFs).  Along with the measured distribution are the
resulting distributions from 100 draws of 25 masses from a
\citet{salpeter1955} IMF and a lower mass limit of 7.5 M$_{\odot}$.
The observed distribution overlaps significantly with many of the
draws, so that a standard IMF is consistent with the data.  A K-S test
shows that half of these draws have a 50\% or higher likelihood of
being drawn from the same parent distribution as the measured masses.
This consistency drops to just 10\% of draws if we assume a minimum
mass of 9.5~M$_{\odot}$.  Thus, we can put a 90\% confidence
upper-limit on least massive stars to produce CCSNe of
$<$9.5~M$_{\odot}$.  Slightly steeper or shallower IMFs, such as those
of \citet{kroupa2001} or \citet{weisz2015}, are similarly consistent
with this sample of SNe.  Thus, while we still do not have any mass
constraint that conclusively rules out a progenitor of mass $\lap$20
M$_{\odot}$, these improved measurements and larger sample do not
require an upper mass cutoff to the supernova progenitor mass
distribution.

Finally, we see no clear association between SN type and progenitor
mass.  The type Ib/c SN appear randomly scattered in mass.  There are
two broad mechanisms for producing stripped massive stars that are
progenitors to SN Ib/c \citep[see][for a good discussion on these
  scenarios]{dessart2011,smith2011}.  In one, because mass loss has a
stiff dependence on luminosity and mass, stars above about 30 solar
masses lose their entire hydrogen envelopes and become Wolf Rayet
stars, the progenitors to SN Ib/c
\citep{langer1994,woosley2002,heger2003}.  In the second, binary
interactions strip the hydrogen envelopes
\citep{woosley1995,yoon2010}.  

Recently, the observed mass-loss rates have been revised downward,
making it more difficult to get the rates of WRs in single-star
evolutionary models \citep{bouret2005}.  More recently,
\citet{smith2011} noted that the rates of stripped envelope SNe are
too high to attribute them to only stars with masses greater than 30
solar masses.  They suggested that the binary evolution scenario is
most consistent with the high rates of stripped-envelope SNe. 

Since binary stripping is more sensitive to orbital parameters and
less sensitive {\bf to the stars' masses}, the binary scenario would produce a
relatively random and broad distribution of progenitor masses.  The
seemingly random and uniform distribution of ages and masses in our
results is at odds with single-star evolutionary models and is most
consistent with the binary evolution scenario.  Taken together, all
these recent results suggest that the progenitors of stripped envelope
SNe are binary systems.  However, we note that the full SFHs of
  these SNe do contain stars of the very youngest ages (see full
  probability distributions).  Thus, while our results prefer type
  Ib/c progenitors of a variety of ages, with this sample of 4 we do
  not rule out high-mass progenitors for these SNe with confidence.

\section{Conclusions}

We have measured resolved stellar photometry around the locations of
12 historic CCSNe and 2 SN impostors.  By fitting color-magnitude
diagrams of this photometry, we have constrained the ages of the SN
progenitor stars.  Using these ages to infer progenitor masses, we
have increased the sample of nearby CCSNe with progenitor mass
constraints.  With our full sample of 25 non-impostor measurements, we
have compared the mass distribution to that of a standard mass
function, such as a \citet{salpeter1955}, \citet{kroupa2001}, or
\citet{weisz2015}.  While the measured masses are somewhat below
expectations from a standard mass function, they are consistent with
such a parent distribution.  Four of our sample are type~Ib/c SNe, and we
find that their progenitors are not the most massive.  Instead, their
ages and masses are uniformly distributed, which suggests that the
progenitors of stripped-envelope SNe are binary systems.

Support for this work was provided by  NASA through grant GO-14786 from the Space Telescope Science Institute, which is operated by the Association of Universities for Research in Astronomy, Incorporated, under NASA contract NAS5-26555.


\begin{deluxetable*}{cccccccc}\tablewidth{6.5in}
\scriptsize
\tablecaption{SNe sample for this study. {\bf Positions taken from the Open
Supernova Catalog (\textrm{\normalfont https://sne.space}), except SN1954A and SN2003gd, which were taken from \citet{Pietra1955} and \citet{maund2009}, respectively.}  Inclinations from the Bright Galaxy Catalog \citep{devaucouleurs1991}.}
\tablehead{
\colhead{SN}  &
\colhead{RA} &
\colhead{DEC} &
\colhead{Galaxy}  &
\colhead{Incl. (deg.)} &
\colhead{Type} &
\colhead{Distance ($Mpc$)} &
\colhead{Previous Measurement M$_{\odot}$}
}
\startdata
SN~1917A  & 308.69542 & 60.12472 & NGC~6946  &  32    & II       &          6.8\tablenotemark{a} & 7.9$^{+6.7}_{-0.5}$ [W14]\\ 
SN~1948B  & 308.83958 & 60.17111 & NGC~6946     &  32 & IIP     &          6.8 & None\\ 
SN~1954A  & 183.9445833 & 36.2630556 &  NGC~4214    & 39   & Ib       &          3.0\tablenotemark{b} & None\\ 
SN~1968D  & 308.74333 & 60.15956 &  NGC~6946    & 32   & II       &          6.8 & None\\ 
SN~1978K  & 49.41083 & -66.55128 &  NGC~1313    &  41  & II       &          4.1\tablenotemark{c} & None\\ 
SN~1980K  & 308.875292 & 60.106597 &  NGC~6946    &  32  & IIL     &          6.8 & $\lap$18 \citep[non-detection][]{thompson1982}\\ 
SN~1985F  & 190.38754 & 41.15164 &  NGC~4618   & 36 & Ib       &          7.9\tablenotemark{d} & None\\
SN~2002ap & 24.099375 & 15.753667 &  NGC~0628  & 24  & Ic pec     &          10.0\tablenotemark{e} & $>$30 \citep[non-detection][]{crockett2007}\\
& & & & & & &  20$-$25 \citep[SN properties][]{mazzali2002}\\ 
SN~2003gd & 24.177708 & 15.738611 &  NGC~0628  & 24  & IIP    &          10.0\tablenotemark{e} & 8$^{+4}_{-2}$ \citep[precursor][]{hendry2005}\\
SN~2004et & 308.85554 & 60.12158 &  NGC~6946    & 32  & IIP     &          6.8 & 56$^{+12}_{-40}$ [W14]\\
& & & & & & & 17$^{+2}_{-2}$ \citep[][]{maund2017}\\ 
& & & & & & & 15$^{+5}_{-2}$ \citep[precursor][]{li2005}\\ 
& & & & & & & 8$^{+5}_{-1}$ \citep[precursor][]{crockett2011}\\ 
SN~2005af & 196.18358 & -49.56661 &  NGC~4945    &  79  & IIP     &          3.6\tablenotemark{f} & 8.5$^{+0.5}_{-0.5}$ [W14]\\ 
SN~2017eaw  & 308.684333 & 60.193306 &  NGC~6946  &  32   & IIP     &          6.8 & $\sim$13 \citep[precursor][preliminary]{vandyk2017}\\
\hline
SN~2002bu & 184.40492 & 45.64650 &  NGC~4242   &  41 & IIn      &          7.9\tablenotemark{d} & Possible Impostor \citep{kochanek2012}\\
SN~2008S  & 308.68896] & 60.09939 &  NGC~6946    &  32  & IIn?     &          6.8 & Possible Impostor \citep{kochanek2012}
\enddata
\tablenotetext{a}{\citet{karachentsev2000}, consistent with \citet{tikhonov2014}}
\tablenotetext{b}{\citealp{dalcanton2009}}
\tablenotetext{c}{\citealp{grise2008}}
\tablenotetext{d}{\citealp{karachentsev2013}}
\tablenotetext{e}{\citealp{jang2014}. This TRGB distance is farther than previously-published distances, and is consistent with our data, which find only 2 main-sequence stars for constraining each of these progenitor masses.}
\tablenotetext{f}{\citealp{monachesi2016}}
\label{sample}
\end{deluxetable*} 


\oddsidemargin -0.5in
\begin{deluxetable*}{llcccccc}\tablewidth{19cm}
\tablecaption{Pointing, Date, Filter, and Exposure Information for HST Observations }
\tabletypesize{\scriptsize}
\tablehead{
\colhead{RA (J2000)}  &
\colhead{Dec (J2000)}  &
\colhead{Date} &
\colhead{Galaxy} &
\colhead{TARGNAME}  &
\colhead{Filter}  &
\colhead{Exp. Time (s)} &
\colhead{Depth (mag)}}
\startdata
308.866178  & 60.1141583 & 2015-10-08,2016-10-26 & NGC~6946 & NGC6946-3\tablenotemark{d} & F438W &   5438 &  28.5\\
308.866178 & 60.1141583 &  2015-10-08,2016-10-26\tablenotemark{e} & NGC~6946 & NGC6946-3\tablenotemark{d} & F606W &   5777 &  28.2\\
308.723173 &  60.1844547 & 2016-10-26 & NGC~6946 & ANY\tablenotemark{a} & F606W &   2430 & 28.5\\
308.723173 &  60.1844547 & 2016-10-26 & NGC~6946 & ANY\tablenotemark{a} & F814W &   2570 &  27.2\\
308.801935 & 60.1712389 & 2016-10-28 & NGC~6946 & NGC6946-1\tablenotemark{b} & F435W &   5610 &  28.1\\
308.801935 & 60.1712389 & 2016-10-28 & NGC~6946 & NGC6946-1\tablenotemark{b} & F606W &   4478 &  27.5\\
308.740159 & 60.1185194 & 2016-11-03 & NGC~6946 & NGC6946-2\tablenotemark{c} & F435W &   5610 &  28.6\\
308.740159 & 60.1185194 & 2016-11-03 & NGC~6946 & NGC6946-2\tablenotemark{c} & F606W &   4488 &  27.9\\
24.1815463 & 15.7285792 & 2016-11-22 & NGC~0628 & SN2003GD & F438W &   2510 & 29.1\\
24.1815463 & 15.7285792 & 2016-11-22 & NGC~0628 & SN2003GD & F606W &   2400 & 27.8\\
24.0861567 & 15.7568500 & 2016-11-22 & NGC~0628 & SN2002AP & F435W &   2300 & 28.3\\
24.0861567 & 15.7568500 & 2016-11-22 & NGC~0628 & SN2002AP & F606W &   2194 & 28.1\\
184.404082 & 45.6430472 & 2016-11-26 & NGC~4242 & SN2002BU & F435W &   2455 & 28.3\\
184.404082 & 45.6430472 & 2016-11-26 & NGC~4242 & SN2002BU & F606W &   2320 & 27.9\\
190.386145 & 41.1467278 & 2016-11-27 & NGC~4618 & SN1985F & F435W &   2410 & 26.6\\
190.386145 & 41.1467278 & 2016-11-27 & NGC~4618 & SN1985F & F606W &   2280 &  26.2\\
183.945776 & 36.2811361 & 2017-01-25 & NGC~4214 & SN1954A & F435W &   1360 & 28.1\\
183.945776 & 36.2811361  & 2017-01-25 & NGC~4214 & SN1954A & F606W &    774 & 27.6
\enddata
\label{obs_table}
\tablenotetext{a}{Parallel exposure that covered the position on SN~2017eaw}
\tablenotetext{b}{SN~1948b and SN~1968D}
\tablenotetext{c}{SN~1917A and SN~2008S}
\tablenotetext{d}{SN~1980K and SN~2004et}
\end{deluxetable*}

\oddsidemargin 0.0in

\begin{deluxetable*}{ccccccc}\tablewidth{14cm}
\tablecaption{SNe Surrounding Population Median Age and Inferred
  Progenitor Mass Constraints}
\tablehead{
\colhead{SN}  &
\colhead{Age (Myr)}  &
\colhead{$+$err} &
\colhead{$-$err} &
\colhead{Mass (M$_{\odot}$)}  &
\colhead{$+$err} &
\colhead{$-$err}
}
\startdata
SN~1917A & 13 & 13 & 1 & 15 & 1 & 5\\
SN~1948B & 28\tablenotemark{a} & 2 & 13 & 10 & 4 & 1\\ 
SN~1954A & 4.5 & 18 & 4.5 &  59 & 1 & 48\\
SN~1968D & 28\tablenotemark{a} & 4 & 14 & 9.6 & 4.2 & 0.6\\
SN~1978K & 34 & 2 & 2 & 8.8 & 0.2 & 0.2\\
SN~1980K & 32\tablenotemark{a} & 1 & 20 & 9.0 & 6.7 & 0.2\\ 
SN~1985F & 21\tablenotemark{a} & 7 & 6 & 11.2 & 2.4 & 1.5\\
SN~2002ap & 18 & 10 & 1 & 12.9 & 0.2 & 2.0\\
SN~2003gd & 47 & 1 & 16 & 7.5 & 1.9 & 0.1\\
SN~2004et & 4.7 & 18 & 1 & 47 & 2 & 36\\
SN~2005af & 5.3 & 15 & 1 & 38 & 1 & 27\\
SN~2017eaw & 33 & 2 & 9 & 8.8 & 2.0 & 0.2\\
\hline
SN~2002bu\tablenotemark{b} & \nodata & \nodata & \nodata & \nodata & \nodata & \nodata\\
SN~2008S & 48 & 2 & 3 & 7.5 & 0.2 & 0.2
\enddata
\tablenotetext{a}{These SNe had probability distributions with multiple significant peaks.  Only the most prominent peak is represented here.  See Table~\ref{age_dist} for the full probability distribution} 
\tablenotetext{b}{This likely impostor had no significant young population within 50 pc.}
\label{mass_table}
\end{deluxetable*}

\begin{deluxetable*}{cccccc}\tablewidth{12cm}
\tablecaption{Age Statistics: Full Table Electronic Only.}
\tabletypesize{\scriptsize}
\tablehead{
\colhead{Supernova}  &
\colhead{Low Age (Myr)}  &
\colhead{High Age (Myr)} &
\colhead{Probability} &
\colhead{$+$err}  &
\colhead{$-$err}  
}
\startdata
SN1917A & 4.0 & 4.5 & 0.000 & 0.017 & 0.000 \\
SN1917A & 4.5 & 5.0 & 0.071 & 0.000 & 0.071 \\
SN1917A & 5.0 & 5.6 & 0.000 & 0.088 & 0.000 \\
SN1917A & 5.6 & 6.3 & 0.000 & 0.091 & 0.000 \\
SN1917A & 6.3 & 7.1 & 0.000 & 0.095 & 0.000 \\
SN1917A & 7.1 & 7.9 & 0.000 & 0.099 & 0.000 \\
SN1917A & 7.9 & 8.9 & 0.000 & 0.109 & 0.000 \\
SN1917A & 8.9 & 10.0 & 0.000 & 0.140 & 0.000 \\
SN1917A & 10.0 & 11.2 & 0.000 & 0.186 & 0.000 \\
SN1917A & 11.2 & 12.6 & 0.000 & 0.309 & 0.000 \\
SN1917A & 12.6 & 14.1 & 0.865 & 0.000 & 0.865 \\
SN1917A & 14.1 & 15.8 & 0.000 & 0.613 & 0.000 \\
SN1917A & 15.8 & 17.8 & 0.000 & 0.588 & 0.000 \\
SN1917A & 17.8 & 20.0 & 0.000 & 0.565 & 0.000 \\
SN1917A & 20.0 & 22.4 & 0.000 & 0.549 & 0.000 \\
SN1917A & 22.4 & 25.1 & 0.000 & 0.506 & 0.000 \\
SN1917A & 25.1 & 28.2 & 0.000 & 0.454 & 0.000 \\
SN1917A & 28.2 & 31.6 & 0.064 & 0.335 & 0.064 \\
SN1917A & 31.6 & 35.5 & 0.000 & 0.308 & 0.000 \\
SN1917A & 35.5 & 39.8 & 0.000 & 0.246 & 0.000 \\
SN1917A & 39.8 & 44.7 & 0.000 & 0.184 & 0.000 \\
SN1917A & 44.7 & 50.1 & 0.000 & 0.118 & 0.000 \\
SN1948B & 4.0 & 4.5 & 0.000 & 0.014 & 0.000 \\
SN1948B & 4.5 & 5.0 & 0.000 & 0.025 & 0.000 \\
SN1948B & 5.0 & 5.6 & 0.017 & 0.032 & 0.017 \\
\nodata & \nodata & \nodata & \nodata & \nodata & \nodata 
\enddata
\label{age_dist}
\end{deluxetable*}

\begin{figure*}
\includegraphics[width=6.75in,valign=m]{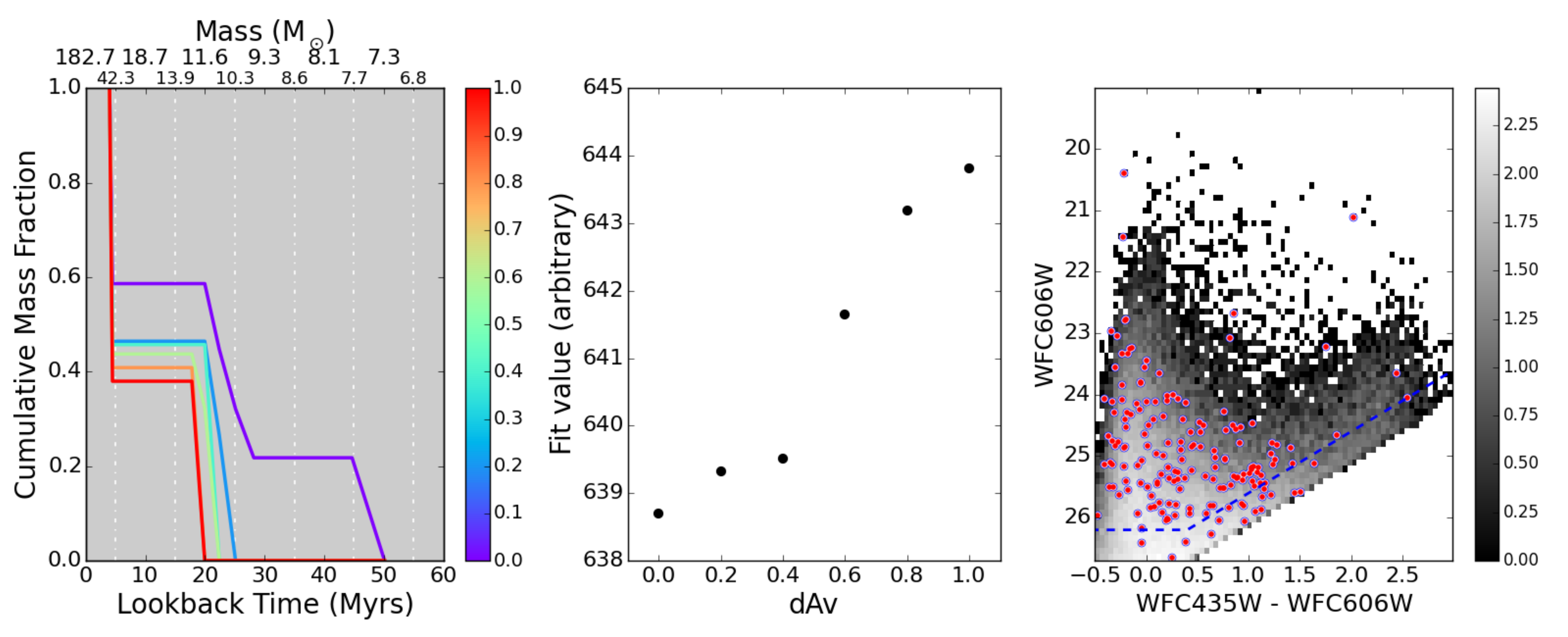}
\includegraphics[width=4.5in,valign=m]{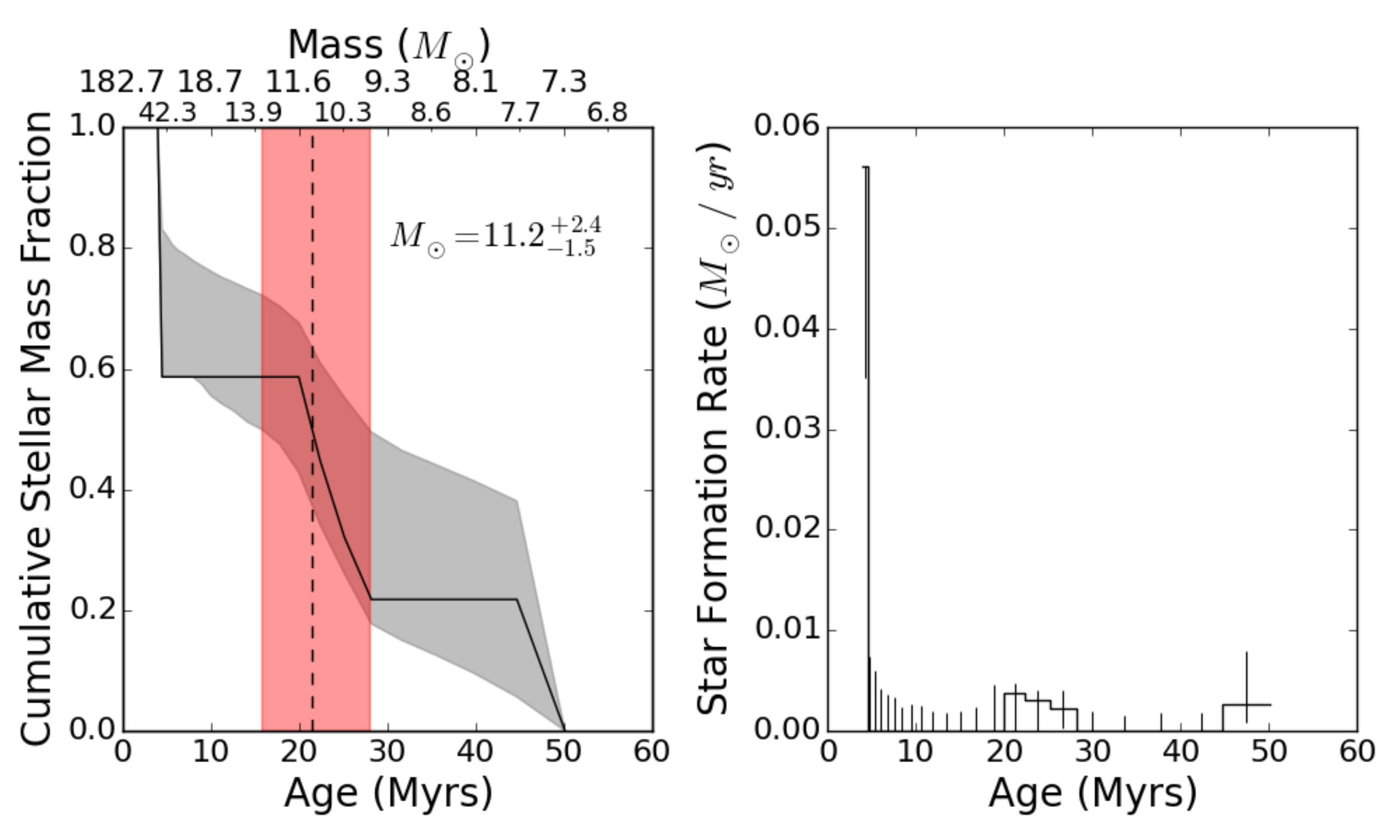}
\includegraphics[width=1.9in,valign=m]{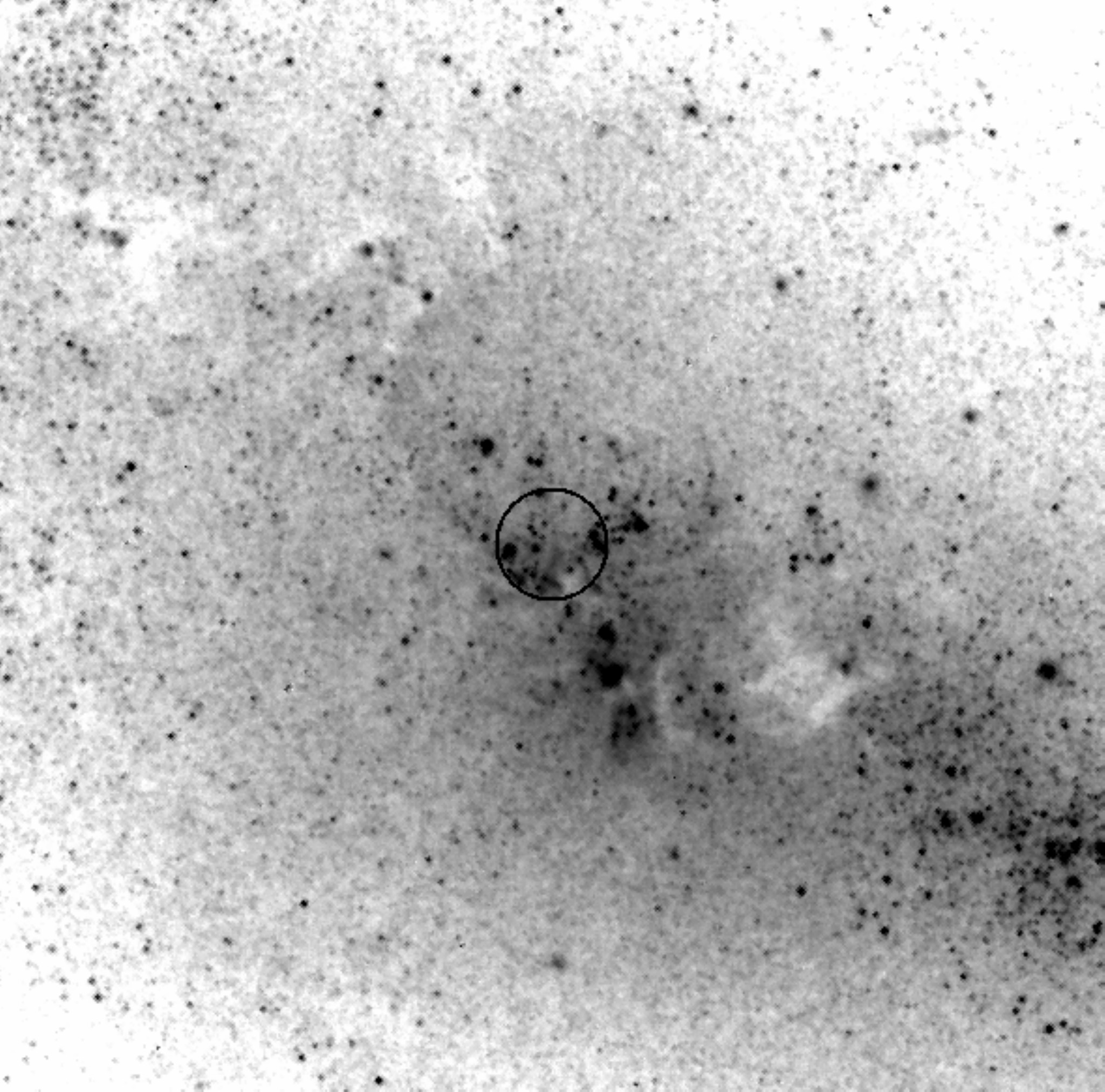}
\caption{Example summary plots for our fitting technique for SN~1985F. These are available for all SN as supplemental material. {\it Top Left:} Cumulative fraction of stars at each age, as measured assuming different amounts of differential reddening.  Each colored line shows the mass distribution measured assuming a different value of dAv.  The color bar indicates the dAv values that correspond to the various colored lines.   {\it Top Middle:} Quality of fit as a function of different amounts of assumed differential reddening.  Lower values indicate a better fit to the data.  {\it Top Right:} CMDs of the data.  Grayscale shows the CMD of the background field (shade denotes log of number of data points in that CMD location).  Outlined red points show the stars inside of the 50pc radius of the SN location.  Blue dashed lines show the region of the CMD included in the fitting.  {\it Bottom Left:} Cumulative fraction of total stellar mass younger than 50 Myr.  The line shows the distribution corresponding to the best fit to the data.  The shaded region shows all distributions consistent with the uncertainties as determined from the hybridMC routine.  The quoted age and uncertainties correspond to the median age of the best fit (dashed line) along with all ages consistent with the median (red shading), considering the uncertainties.  However, this single value with error does not always accurately capture the age distribution, so we include the full probability distribution in Table~\ref{age_dist}. {\it Bottom Middle:}  The differential SFH, showing the rates and uncertainties in each age bin.  This view is difficult to interpret directly because there are large covariances between adjacent age bins. {\it Bottom Right:} 25$''{\times}25''$ {\it HST} image of the region surrounding the SN location.  The orientation is North-up, East-left, and the 50 pc circle from which we took resolved stellar photometry for fitting is marked.}
\label{example1}
\end{figure*}

\begin{figure*}
\includegraphics[width=6.75in,valign=m]{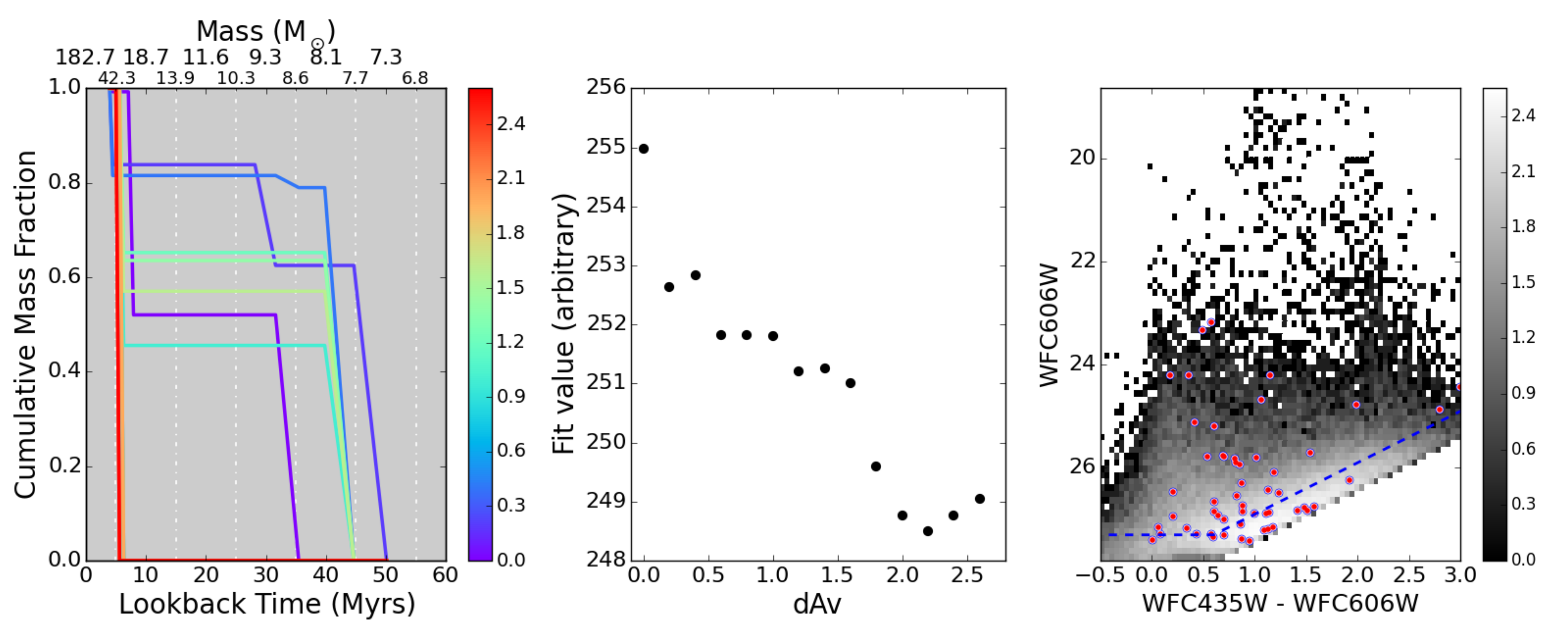}
\includegraphics[width=4.5in,valign=m]{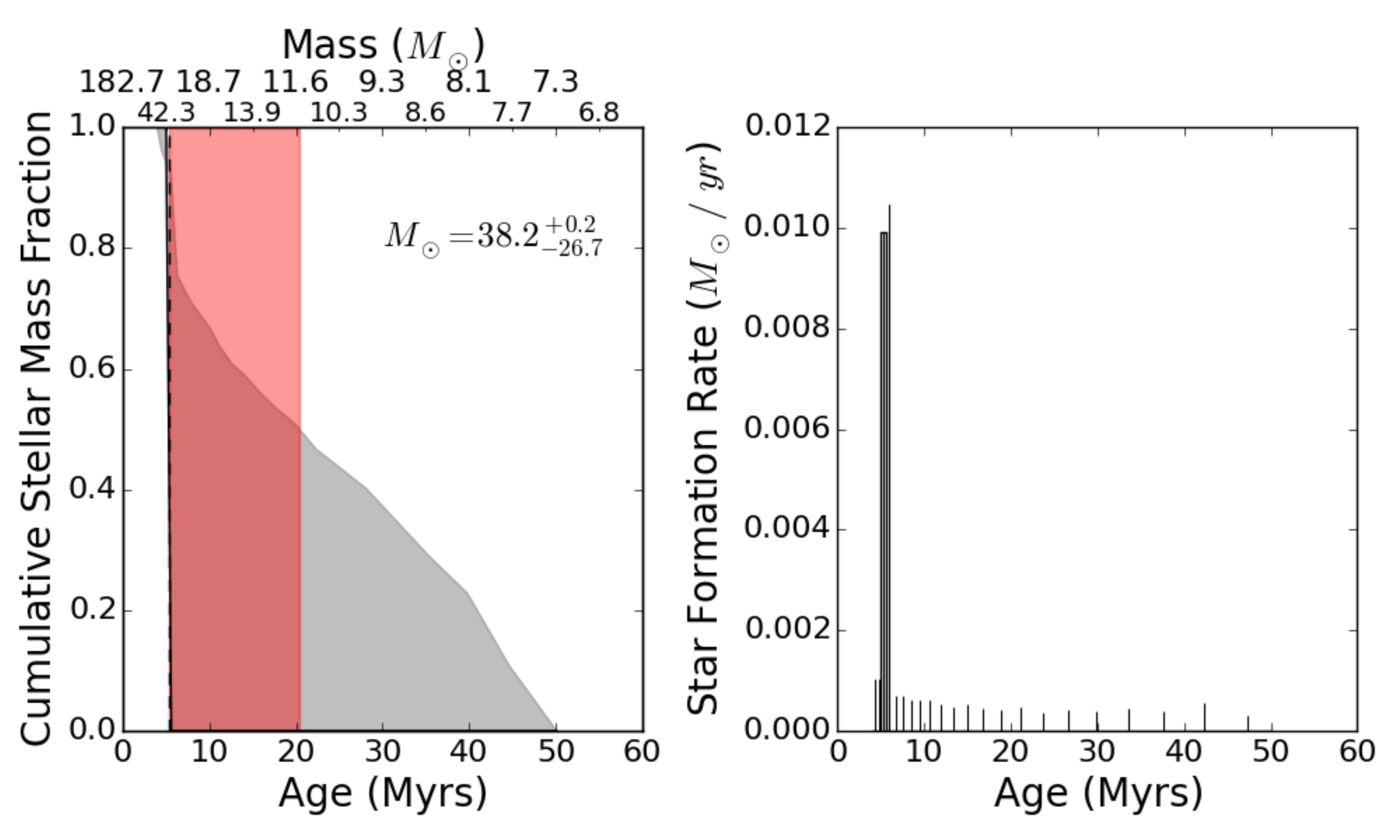}
\includegraphics[width=1.9in,valign=m]{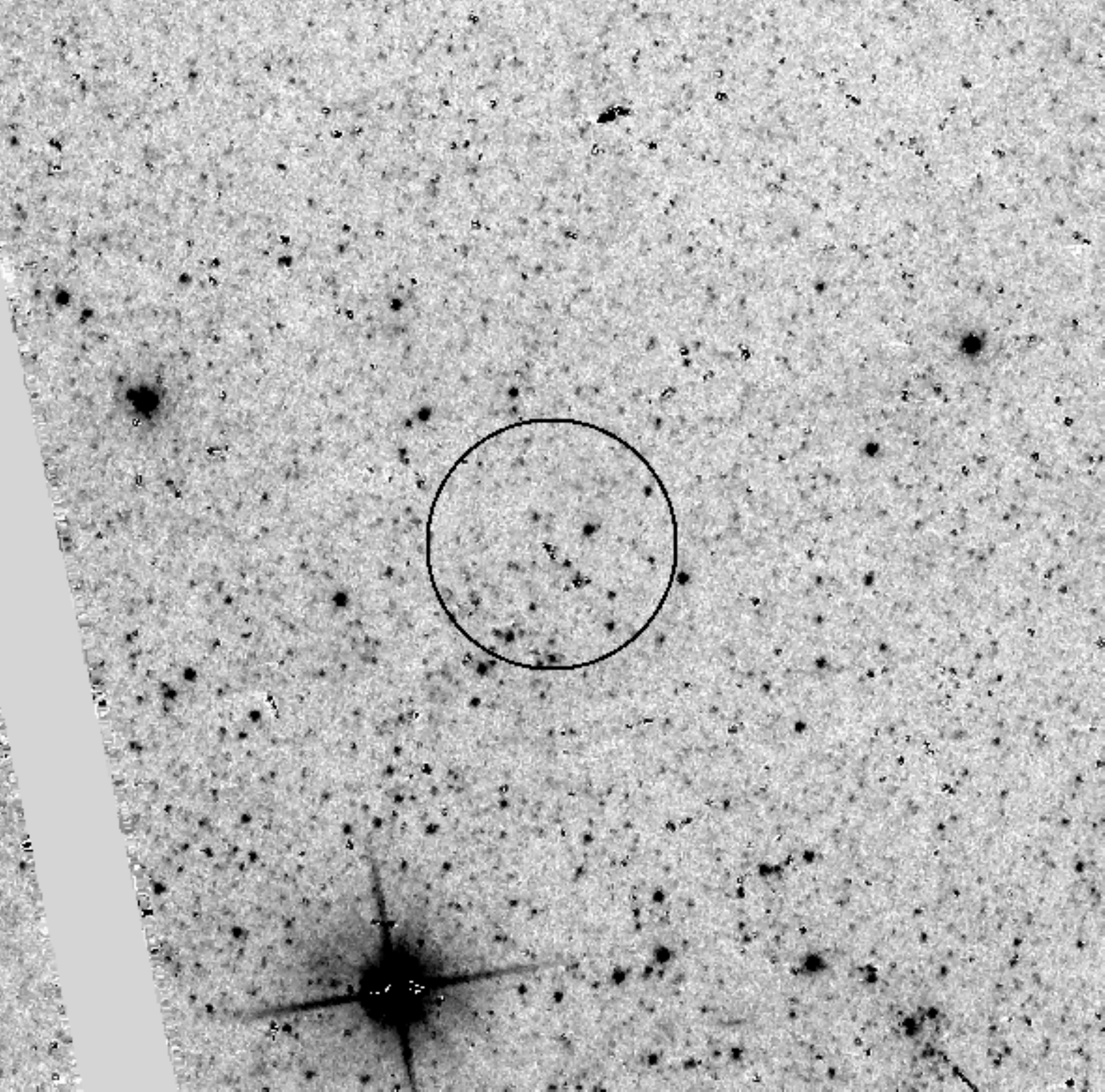}
\caption{Same as Figure~\ref{example1} summary but for SN~2005af.}
\label{example2}
\end{figure*}

\begin{figure*}
\begin{center}
\includegraphics[width=4.0in]{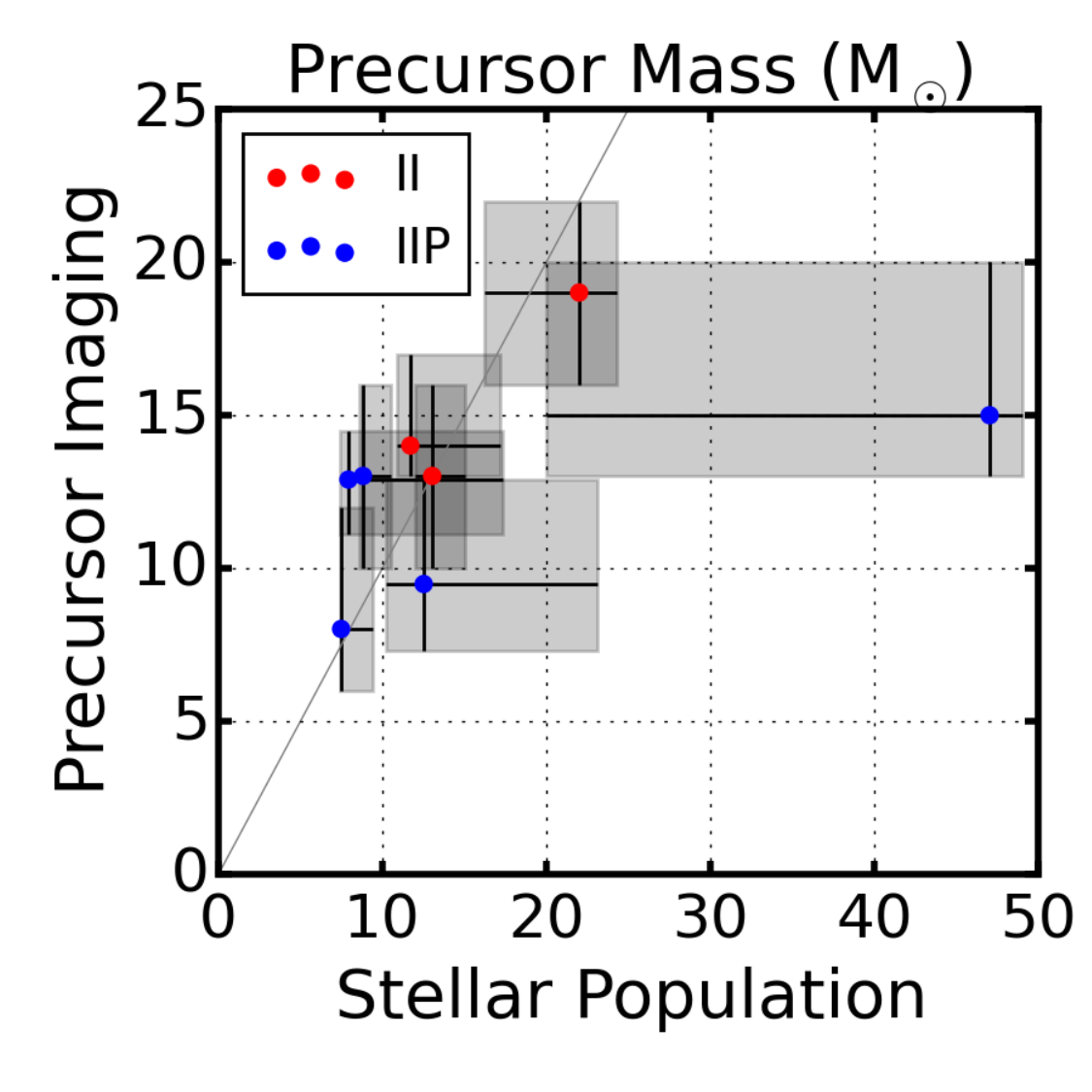}
\caption{Comparison of progenitor masses measured with direct precursor SED fitting vs those from our stellar population analysis where both measurements are available (taking measurements from here, \citealp{murphy11a}, and W14).  There is good agreement within the uncertainties in all cases. The farthest outlier is SN~2004et, which lies in a region with a young population with a large uncertainty in age, but best modeled by a very young age.}
\label{compare}
\end{center}
\end{figure*}

\begin{figure*}
\begin{center}
\includegraphics[width=5.0in]{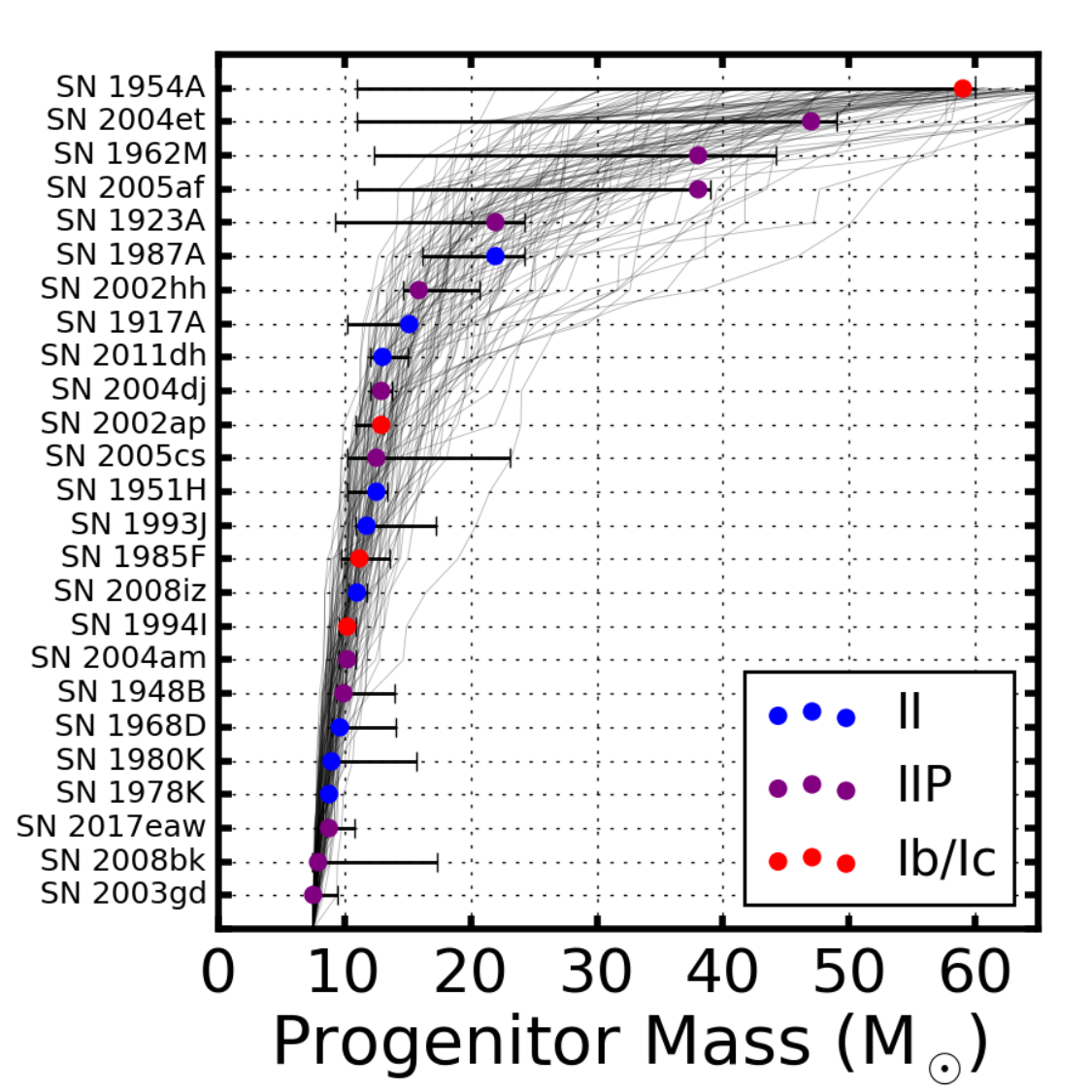}
\caption{The masses from the median ages, and their uncertainties, ranked by mass are shown with the points and error bars.  Blue points show Type II SNe, and red points show Type Ib/c SNe.  Likely impostors have not been included.  The Type Ib/c SNe are not concentrated at the highest masses but randomly distributed in age and mass.  This suggests that progenitors of stripped-envelope SNe are likely binary stars. Gray lines show draws of 25 stars from a \citet{salpeter1955} mass function without errors and a minimum mass of 7.5~M$_{\odot}$.  The standard mass function, as well as a slightly steeper mass function such as that of \citet{weisz2015}, are fully consistent with the data.}
\label{mass_dist}
\end{center}
\end{figure*}

\end{document}